\documentclass[journal=jacsat,manuscript=article]{achemso}

\usepackage[version=3]{mhchem} % Formula subscripts using \ce{}

\author{Elnaz Zyaee}
\affiliation{Institute of Applied Physics, University of Bern}
\email{elnaz.zyaee@unibe.ch}

\author{Vladislav Slama}
\affiliation{Laboratory of Computational Chemistry and Biochemistry, École Polytechnique Fédérale de Lausanne (EPFL)}
\alsoaffiliation{Faculty of Mathematics and Physics, Charles University}%, Ke Karlovu 3, 121 16, Prague, Czech Republic}

\author{Seyyed Jabbar Mousavi}
\affiliation{Institute of Applied Physics, University of Bern}
\author{David Rohrbach}
\affiliation{Institute of Applied Physics, University of Bern}

\author{Ursula Rothlisberger}
\affiliation{Laboratory of Computational Chemistry and Biochemistry, École Polytechnique Fédérale de Lausanne (EPFL)}

\author{Thomas Feurer}
\affiliation{Institute of Applied Physics, University of Bern}

\title[An \textsf{achemso} demo]
  {Time-resolved THz Stark spectroscopy of molecules in water}
\abbreviations{IR,NMR,UV}
\keywords{American Chemical Society, \LaTeX}

\begin{document}

\begin{abstract}
Stark spectroscopy is a powerful method for probing molecular dipole moment changes, charge transfer dynamics, and polarizability under applied electric fields. Time-Resolved Terahertz Stark Spectroscopy (TRTSS), which employs intense single-cycle terahertz (THz) pulses to induce transient Stark shifts, overcomes key limitations of conventional approaches. Unlike static or low-frequency fields, THz pulses oscillate much faster than typical molecular rotation times, effectively preventing dipole reorientation and enabling measurements in solutions at ambient conditions.

Here, we extend TRTSS to molecules dissolved in water, the most important polar solvent for chemical and biological systems and report the first demonstration of Stark spectroscopy in water at room temperature. Using Malachite Green and Methyl Orange as model systems, we observe clear THz-induced spectral modulations, demonstrating that TRTSS can successfully reveal THz Stark responses even in highly polar, hydrogen-bonded environments. 
Measured signals exhibit a combination of linear (dipole-driven) and quadratic (polarizability-driven) Stark effects in both systems, consistent with time-dependent density functional theory (TD-DFT) calculations. Comparison with TD-DFT further suggests that conformational effects can influence the extracted Stark parameters in solvated molecules.
\end{abstract}

\newpage

\section{Introduction}
Understanding how molecules respond to external electric fields provides fundamental insight into their electronic structure, charge distribution, and interactions with their environment. Stark spectroscopy has long been an essential tool for probing these properties by analyzing electric-field-induced spectral shifts, revealing information about dipole moments, polarizabilities, and charge transfer processes \cite{Brunschwig1998, Bublitz1997, Iimori2016, Liptay1969, Pein2019terahertz}. However, conventional Stark spectroscopy typically relies on low-frequency (kHz) electric fields applied through electrodes, necessitating cryogenic immobilization of samples to avoid rotational and alignment effects\cite{Bublitz1997Stark,Andrews2000VSEI,Andrews2000Cryostat}. This limitation restricts the range of solvents that can be used, often excluding biologically and chemically relevant polar environments such as water, where solvation effects play a crucial role in molecular behavior\cite{Bublitz1997Stark,Andrews2000VSEI}.

Recently, terahertz (THz) Stark spectroscopy has emerged as a powerful alternative, utilizing ultrafast, phase-stable THz pulses to transiently induce Stark shifts without requiring sample freezing \cite{Pein2019terahertz, Keiber2016electro, Knorr2017phase, kang2024, Singh2023, Zhang2024}. In our previous work, we demonstrated that THz Stark spectroscopy enables precise measurements of dipole moment changes and polarizabilities in non-polar and weakly polar solvents under ambient conditions \cite{kang2024}. Around the same time, Singh et al. reported the THz Stark spectroscopy of a prototypical organic dye, Betaine-30 (B-30), in two aprotic polar solvents: chloroform (CHCl$_3$) and dimethylsulfoxide (C$_2$H$_6$SO, DMSO)\cite{Singh2023}. More recently, they also employed THz Stark spectroscopy to monitor the electric dipole changes of wild-type bacteriorhodopsin (BR) and a BR D85T mutant upon electronic excitation\cite{Zhang2024}. However, the applicability of this technique to water, a highly polar solvent, has not yet been explored.

Extending TRTSS to water presents fundamental challenges. First, the frequency-dependent complex permittivity of water in the THz range modifies the electric field experienced by solvated molecules and must therefore be considered. \cite{Ellison2007PermittivityWater,Liebe1991WaterPermittivity,Aubret2019}
%when determining the effective local field.  
Second, strong THz absorption by water, due to intermolecular hydrogen bonding and dipolar relaxation raises concerns about effective field penetration and whether the induced Stark shifts remain detectable \cite{Kindt1996FarInfrared,10.1063/1.5047659}. 
Finally, specific solute-solvent interactions, including hydrogen bonding and local electrostatic effects, may modify the electronic structure and THz Stark response of chromophores in water, complicating the interpretation of the extracted molecular parameters.\cite{Bublitz1997Stark,annurev:/content/journals/10.1146/annurev-physchem-052516-045011}

In this study, we demonstrate THz Stark spectroscopy is feasible in water and provide valuable molecular insights previously inaccessible with conventional Stark methods. Using Malachite Green and Methyl Orange as model systems, we show that clear THz Stark signals can be detected even in water, enabling precise measurements of molecular parameters such as dipole moment changes and polarizabilities in aqueous environment. The experimentally extracted values are compared with equilibrium-solvated TD-DFT calculations.
This comparison further allows to assess how molecular structure and conformational fluctuations influence the resulting Stark parameters in water. 

\section{Results and discussion}
\subsection{Observation of THz Stark Dynamics in Water}
\begin{figure} [H] %[ht!]
    \centering
    \includegraphics[width=0.8\columnwidth]{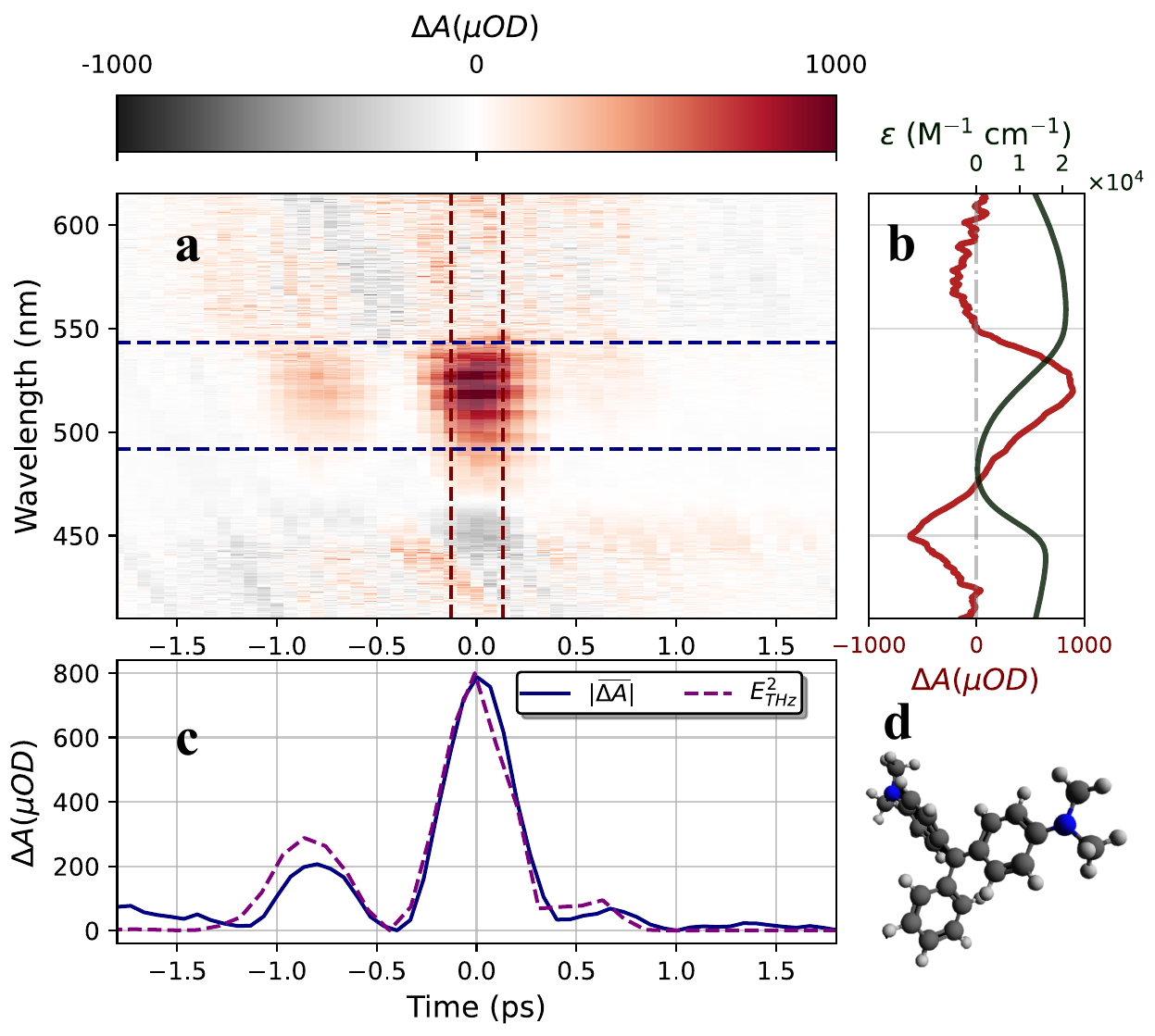}
    \caption{\label{fig:THzStarkMalachiteGreen} 
    \textbf{THz Stark response of Malachite Green in the parallel polarization geometry.}
    \textbf{a} False-color map of the measured change in absorption spectrum as a function of wavelength and probe delay relative to the THz pulse.  
    \textbf{b} Time-averaged Stark spectrum obtained over the time window indicated by the red dashed lines in \textbf{a} (red curve), shown together with the steady-state molar absorption coefficient \(\varepsilon(\lambda)\) (green curve).   
    \textbf{c} Temporal evolution of the spectrally integrated THz Stark signal (blue curve), averaged across the spectral range indicated by the blue dashed lines in \textbf{a}, compared to the squared THz electric field profile $E_\mathrm{THz}^2$ (purple dashed curve).  
    \textbf{d} Chemical structure formula of Malachite Green.
}
\end{figure}

\begin{figure} [H] %[ht!]
    \centering
    \includegraphics[width=0.8\columnwidth]{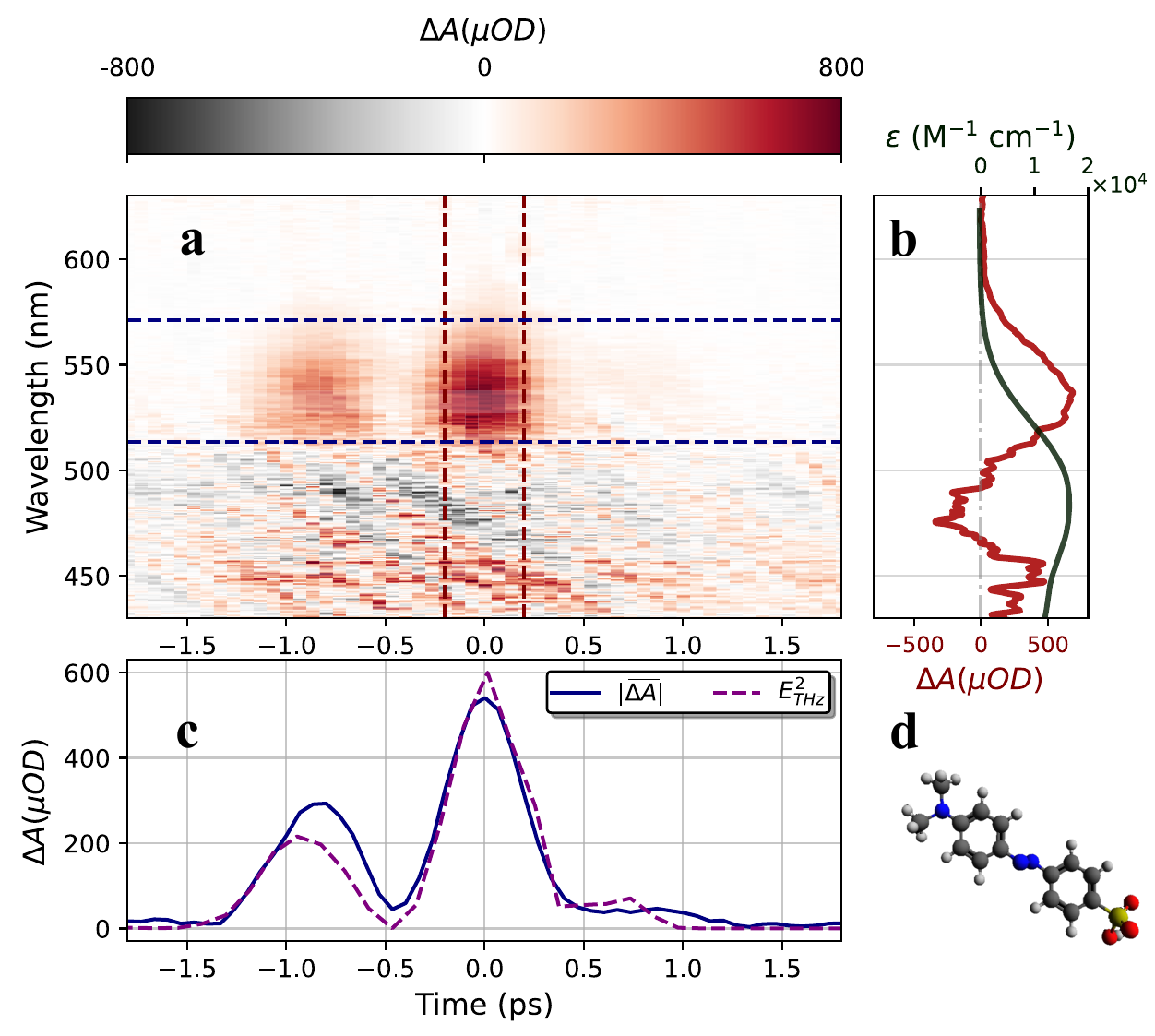 }
    \caption{\label{fig:THzStarkMethylOrange} 
    \textbf{THz Stark response of Methyl Orange in the parallel polarization geometry.}
    \textbf{a} False-color map of the measured change in absorption spectrum as a function of wavelength and probe delay relative to the THz pulse.  
    \textbf{b} Time-averaged Stark spectrum obtained over the time window indicated by the red dashed lines in \textbf{a} (red curve), shown together with the steady-state molar absorption coefficient \(\varepsilon(\lambda)\) (green curve).
    \textbf{c} Temporal evolution of the spectrally integrated THz Stark signal (blue curve), averaged across the spectral range indicated by the blue dashed lines in \textbf{a}, compared to the squared THz electric field profile $E_\mathrm{THz}^2$ (purple dashed curve).  
    \textbf{d} Chemical structure of Methyl Orange.}
\end{figure}

This study presents the THz-induced Stark response of two dyes dissolved in water, Malachite Green (Figure~\ref{fig:THzStarkMalachiteGreen}) and Methyl Orange (Figure~\ref{fig:THzStarkMethylOrange}). To isolate the molecular response of each dye and remove the nonresonant Kerr/electrostriction background, a reference measurement of pure water was performed under identical conditions, including cuvette, path length, polarization geometry, and THz energy (see Supporting Information, Figure~\ref{fig:Water spectroscopy}). 
The pure-water response was then subtracted from the dye-solution data. For both dyes, this subtraction changed the RMS magnitude of the extracted THz Stark response by up to approximately \(1\%\) within the selected spectral and temporal windows.

In the TRTSS experiment, an intense single-cycle THz pulse was spatially overlapped with a broadband white-light probe pulse at the sample. The THz pulse induced transient changes in the molecular absorption spectrum, which were recorded in transmission as a function of probe wavelength and THz--probe delay. Measurements were performed with the probe polarization oriented either parallel or perpendicular to the THz electric-field polarization.

Figure~\ref{fig:THzStarkMalachiteGreen} presents the THz Stark response of Malachite Green measured in the parallel polarization geometry. Figure~\ref{fig:THzStarkMalachiteGreen}\textbf{a} shows a false-color map of the THz Stark response, expressed as \(\Delta A(\lambda,t)\), as a function of probe wavelength and THz--probe delay. The response is confined to the 440--550~nm spectral region and reaches its maximum amplitude at \(t=0\), defined by temporal overlap of the THz and probe pulses.

Figure~\ref{fig:THzStarkMalachiteGreen}\textbf{b} shows the time-averaged THz-induced spectral response of Malachite Green over the time window indicated by the red dashed lines in panel~\textbf{a}. The steady-state molar absorption coefficient is shown on the same plot. The modulation is localized around the structured absorption band, and changes sign across the band. Around the lower-energy absorption feature near 550~nm, the THz Stark spectrum shows a weak zero crossing close to the steady-state absorption maximum, consistent with a predominantly field-induced shift-like contribution in this spectral region. At higher energies, the strongest THz Stark response occurs near the absorption maximum around 450~nm, indicating a field-induced broadening-like contribution. Because the two contributions overlap, it can be assigned neither to a purely shift- or broadening-type response. The overall response therefore reflects a mixed Stark modulation of the structured absorption band.

Figure~\ref{fig:THzStarkMalachiteGreen}\textbf{c} depicts the time-domain evolution of the THz Stark signal, obtained by averaging over the spectral region marked by the blue dashed lines in panel~\textbf{a}. The THz Stark response follows the squared THz electric-field profile, \(E_\mathrm{THz}^2\) (purple dashed curve), and is confined to the duration of the applied field, indicating that it is driven by the instantaneous THz field intensity. This demonstrates that the solvent does not induce a memory effect in the molecular response, especially no noticeable effect from reorientation of the water molecules.
Finally, Figure~\ref{fig:THzStarkMalachiteGreen}\textbf{d} displays the chemical structure of Malachite Green. 

Figure~\ref{fig:THzStarkMethylOrange} presents the THz-induced response of Methyl Orange in water, measured with the probe polarization parallel to the THz electric field. The false-color plot in Figure~\ref{fig:THzStarkMethylOrange}\textbf{a} shows the time-dependent change in absorption, \(\Delta A(\lambda,t)\), with the strongest modulation observed in the 500--580~nm range around temporal overlap. This indicates a THz-induced response of Methyl Orange in water and further demonstrates the applicability of TRTSS in highly polar solvents.

In Figure~\ref{fig:THzStarkMethylOrange}\textbf{b}, the THz Stark response is centered around the absorption band, with a pronounced negative feature near the band maximum and positive contributions on both spectral sides. The overall lineshape is therefore consistent with predominantly field-induced broadening of the absorption band, while the imperfect symmetry of the extrema around the absorption maximum indicates an additional smaller shift-like contribution.

Figure~\ref{fig:THzStarkMethylOrange}\textbf{c} shows the corresponding time trace, obtained by averaging \(\Delta A\) over the wavelength range marked by the blue dashed lines in panel~\textbf{a}. As observed for Malachite Green, the THz Stark signal follows the squared THz electric-field profile, \(E_\mathrm{THz}^2\), supporting its assignment to a field-driven Stark response. Figure~\ref{fig:THzStarkMethylOrange}\textbf{d} shows the molecular structure of Methyl Orange.

Together, the Malachite Green and Methyl Orange measurements demonstrate that molecular THz Stark responses can be detected in water despite the strong absorption and nonresonant background of the solvent. In both cases, the spectral modulation is confined to the molecular absorption region and changes sign across the band, while the temporal response tracks \(E_\mathrm{THz}^2\), supporting its assignment to a field-driven Stark response.

This measurement is challenging because water attenuates THz radiation, modifies the local electric field through its frequency-dependent dielectric response, and produces a nonresonant birefringence/Kerr-type background that overlaps temporally with the molecular signal. After subtraction of the pure-water response, however, the remaining dye-specific THz Stark spectra retain the derivative-like spectral shape expected for molecular Stark spectra. This result shows that molecular Stark contributions can be separated from the water background and extends TRTSS from weakly polar organic solvents to a regime where dielectric screening, THz attenuation, and specific solvation effects are all significant.

\subsection{Molecular Stark parameters and comparison with TD-DFT}

Having established the presence of THz-induced molecular responses in water, we quantified the Stark spectra using the Liptay electroabsorption formalism. The measured spectral modulation was described as a linear combination of the first and second derivatives of the steady-state absorption spectrum, following the standard formalism for isotropic samples \cite{Liptay1969,Bublitz1997Stark,JALVISTE200730}. This analysis enables extraction of molecular Stark parameters, including the magnitude of the dipole moment change, \(|\Delta\boldsymbol{\mu}|\), the angle \(\zeta\) between the dipole-change vector \(\Delta\boldsymbol{\mu}\) and the transition dipole moment \(\boldsymbol{m}\), and the average change in polarizability, \(\mathrm{Tr}(\Delta\boldsymbol{\alpha})\).

For an isotropic solution, the Liptay formalism also predicts a characteristic polarization dependence. Spectra measured with probe polarization parallel and perpendicular to the THz field should be described by the same derivative basis, while their amplitudes differ through orientational averaging factors \cite{Liptay1969,Bublitz1997Stark,JALVISTE200730}. We therefore measured both polarization geometries as an internal consistency check of the molecular Stark assignment. The detailed polarization comparison, including the best-scaled parallel and perpendicular spectra, is provided in the Supporting Information (see SI, Figure~\ref{fig:Polarization anisotropy}).

\begin{figure} [H] %[ht!]
    \centering
    \includegraphics[width=1\columnwidth]{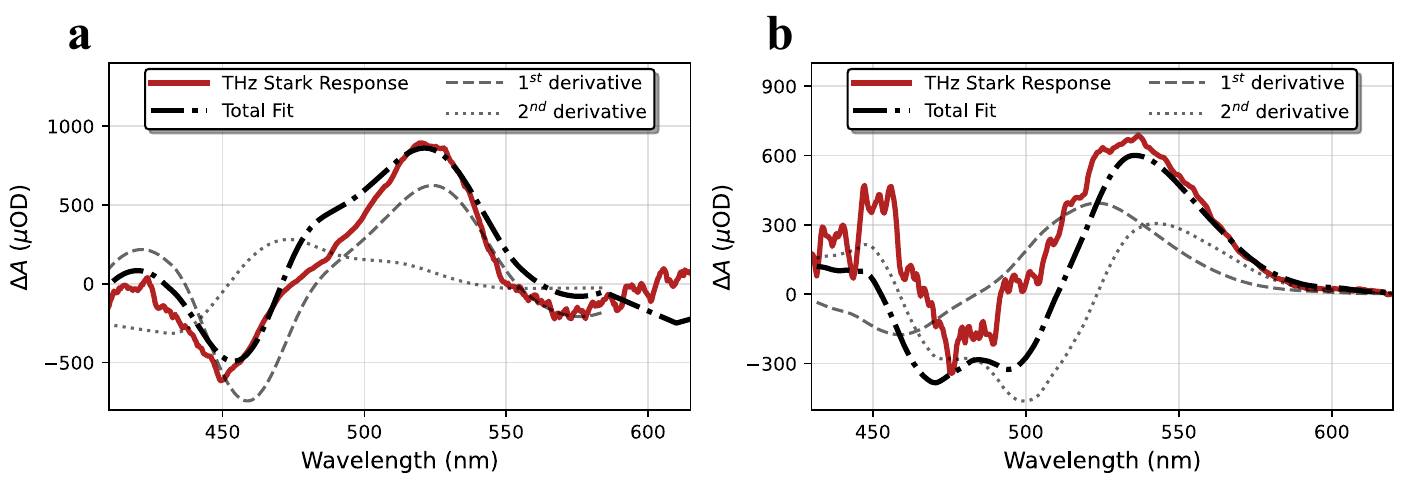}
    \caption{\label{fig:MGMO_AttenuationCoefficent}
    \textbf{Liptay analysis of the THz Stark spectra.}
    Time-averaged THz Stark spectra of \textbf{a} Malachite Green and \textbf{b} Methyl Orange in water, shown together with the Liptay derivative decomposition. The measured Stark spectra are shown in red, the total fit in black, and the first- and second-derivative contributions of the ground-state absorption spectrum as gray dashed and dotted curves, respectively. Both molecules are described by a mixed THz Stark response.
    }
\end{figure}

Figure~\ref{fig:MGMO_AttenuationCoefficent} shows the Liptay derivative decomposition for the TRTSS spectra of Malachite Green and Methyl Orange in water. For Malachite Green (Fig.~\ref{fig:MGMO_AttenuationCoefficent}\textbf{a}), the first-derivative contribution dominates the fit, while a smaller second-derivative contribution is required to reproduce the full spectral shape. For Methyl Orange (Fig.~\ref{fig:MGMO_AttenuationCoefficent}\textbf{b}), the spectrum also requires a mixed derivative description, but the second-derivative contribution is more pronounced and plays a major role in reproducing the asymmetric modulation around the absorption band. These results show that both molecules exhibit mixed Stark behavior, with different relative contributions from dipole-moment and polarizability changes. 

Using the fitted derivative components, we extracted the molecular Stark parameters summarized in Table~\ref{tab:ExtractedParameterMGMO}. The experimentally extracted values are compared with equilibrium-solvated TD-DFT calculations based on optimized geometries. Additional TD-DFT calculations on geometries extracted from classical molecular dynamics simulations were used to assess the effect of thermal geometrical fluctuations on the calculated Stark parameters.
\begin{table}[H] %[ht!]
\caption{\label{tab:ExtractedParameterMGMO} Comparison of relevant molecular parameters as calculated via equilibrium-solvated TD-DFT and THz Stark spectroscopy.}
\begin{tabular}{ |p{4cm}|p{5.5cm}|p{5.5cm}| }
\hline
Parameter & THz-Stark & equilibrium-solvated TD-DFT  \\
\hline
\multicolumn{3}{|c|}{Malachite Green} \\
\hline
$|\Delta\boldsymbol{\mu}|$ (D)   & $5.6 \pm 1.2 $  &  4.2  \\
$\zeta$ ($^\circ$)   & $36.32\pm 1.1$ &  90  \\
Tr($\Delta\boldsymbol{\alpha}$) (\AA$^3$)   & $-57.3 \pm 15.4 $  & 32   \\
%$m\Delta\alpha m/m^2$ (\AA$^3$)   & 32   &   $ -37.7 \pm 8.9 $ \\
\hline
\hline
\multicolumn{3}{|c|}{Methyl Orange} \\
\hline
$|\Delta\boldsymbol{\mu}|$ (D)   & $6.2\pm 1.8 $   &  8.7 \\
$\zeta$ ($^\circ$)   & $24.76\pm 2.1$ &  9  \\
Tr($\Delta\boldsymbol{\alpha}$) (\AA$^3$)   & $115.3 \pm 17.2 $ &  92  \\
%$m\Delta\alpha m/m^2$ (\AA$^3$)   &  91 &   $79.9  \pm 12.5 $  \\
\hline
\end{tabular}
\end{table}
 The validity of the TD-DFT approach for simultaneous treatment of both charge-transfer (CT) and locally excited states was checked by comparison with gas phase CC2 calculations (see SI). The experimentally extracted \(\zeta\) angles differ from the equilibrium-solvated TD-DFT values, particularly for Malachite Green. A comparable limitation was encountered in our previous TRTSS study, where \(\zeta\) could not be determined reliably from the data.\cite{kang2024} The present comparison therefore indicates that \(\zeta\) is especially sensitive to molecular geometry, conformational averaging, and the polarization dependence of the THz Stark response. In the calculations with equilibrium solvation, only optimized geometries are considered, whereas thermally induced distortions can modify both the transition dipole moment and the direction of \(\Delta\mu\).

The pronounced sensitivity of the mutual angle $\zeta$ to the molecular structure arises from the strong structural dependence of the spatial localization of the HOMO and LUMO molecular orbitals, which determines both the resulting transition dipole moment associated with the locally excited contribution and the charge redistribution associated with the CT contribution to the lowest excited state. While the transition dipole moment depends on the spatial overlap of the HOMO and LUMO orbitals, the excited-state dipole moment associated with the CT contribution depends on their spatial separation.
For example, in Malachite Green, geometrical fluctuations can localize the HOMO and LUMO orbitals on one of the dimethylaminophenyl groups, resulting in a transition dipole moment oriented from the group toward the center of the molecule and charge transfer from one dimethylaminophenyl group to the other or to the phenyl group. This behavior is demonstrated by TD-DFT calculations of the electronic properties for molecular geometries extracted from room temperature classical molecular dynamics (MD) simulations (see SI, Figure~\ref{fig:MG_MD_geometry}). Such geometrical distortions reduce the $\zeta$ angle from 90$^\circ$ in the optimized ground-state geometry to 10$^\circ$ and 24$^\circ$ for the first and second excited states, respectively. Thermally induced localization of the molecular orbitals also affects the change in molecular polarizability, for example, $Tr(\Delta \alpha) = -45 \,  \AA^3$ for the second excited state, which is close to the experimental value. In contrast, significantly smaller effects of thermal geometrical fluctuations are expected for Methyl Orange because of stabilization of the planar structure by $\pi$-electron delocalization across the two aromatic rings and the azo \((-N=N-)\) linkage. We note that, compared with quantum-mechanical descriptions of the potential energy surface, classical MD simulations may overestimate structural fluctuations owing to the approximations inherent in empirical, non-polarizable force fields. Therefore, the present analysis should be regarded as qualitative rather than quantitative.

Together, the derivative-like spectral response, field-squared temporal evolution, polarization behavior, and TD-DFT comparison support the assignment of the observed signals to molecular THz Stark responses in water. The quantitative extraction of molecular parameters further validates TRTSS as a probe of solvated molecular systems under ambient conditions.

\section{Conclusion}
 
In this study, we have demonstrated for the first time that TRTSS can be successfully applied to highly polar solvents such as water under ambient conditions. Our findings confirm that solvent polarity and hydrogen bonding do not fundamentally limit the ability to measure Stark shifts in solution, significantly expanding the applicability of TRTSS beyond non-polar solvents.

By investigating Malachite Green and Methyl Orange, we observed clear THz Stark signals that exhibit a combination of linear (dipole-driven) and quadratic (polarizability-driven) Stark effects, suggesting the mixed THz Stark response of these molecules. The comparison with TD-DFT calculations using equilibrium solvation, complemented by calculations on thermally distorted geometries, provides further support for the observed mixed Stark behavior. It also demonstrates the capability of the TD-DFT approach employing the $\omega$B97XD range-separated functional to describe molecular excitations with mixed locally excited and charge-transfer (CT) character. 
The \textit{ab initio} calculations help connect the observed spectral features to changes in molecular electronic structure and identify involved molecular orbitals. Based on this analysis, we further demonstrated the different sensitivities of the molecules to thermal fluctuations, arising from the interplay between the steric effects and $\pi$-electron delocalization.

These results mark a major advancement for THz Stark spectroscopy, an invaluable tool for determining the electronic properties of molecules, proving that the technique can be applied to liquid-phase samples without freezing, even in highly polar solvents. Moreover, our findings establish TRTSS as a versatile and solvent-compatible spectroscopic tool, paving the way for new explorations in biophysics, ultrafast electrochemistry, and solution-phase quantum dynamics.
Importantly, the combined TRTSS, Liptay analysis, and equilibrium-solvated TD-DFT approach provides a quantitative framework for connecting experimentally observed THz Stark responses with underlying excited-state electronic structure.

Future work will explore more complex molecular systems, including biological molecules and electrochemical interfaces, where solvent interactions play a crucial role. Additionally, expanding the method to higher-field THz pulses Stark spectroscopy could provide deeper insights into nonlinear Stark effects and transient electronic structures in solution-phase dynamics.

\newpage

%%%%%%%%%%%%%%%%%%%%%%%%%%%%%%%%%%%%%%%%%%%%%%%%%%%%%%%%%%%%%%%%%%%%%
\newpage

\section{Data Availability}
The data supporting the findings of this study are available from the authors upon request.
\begin{acknowledgement}
This work was supported by the Swiss National Science Foundation (SNSF) through the Sinergia grant CRSII5-213533 and grants 200021-204053 and 200020-185092. The authors also acknowledge the Swiss National Computing Centre (CSCS) for providing computational resources.

\end{acknowledgement}

\newpage

\begin{suppinfo}

\subsubsection{THz Stark Spectroscopy Setup}
Time-Resolved THz Stark Spectroscopy (TRTSS) was performed using a single-cycle terahertz (THz) pulse to induce transient Stark shifts in molecular absorption spectra. Figure \ref{fig:Setup} provides an overview of the experimental setup. The THz pulses were generated via optical rectification in a LiNbO$_3$ crystal using a tilted-pulse-front scheme. A synchronized femtosecond supercontinuum probe pulse, generated through white-light continuum generation in a CaF$_2$ crystal, was used to monitor Stark-induced absorption changes. The probe and THz pulses were spatially overlapped and temporally delayed to scan the THz Stark response as a function of time delay. The transmitted probe spectra were detected using a 1024-pixel CMOS array, with a reference line to enable shot-to-shot normalization and enhance the signal to noise ratio.
\begin{figure} [ht!]
    \centering
    \includegraphics[width=0.8\linewidth]{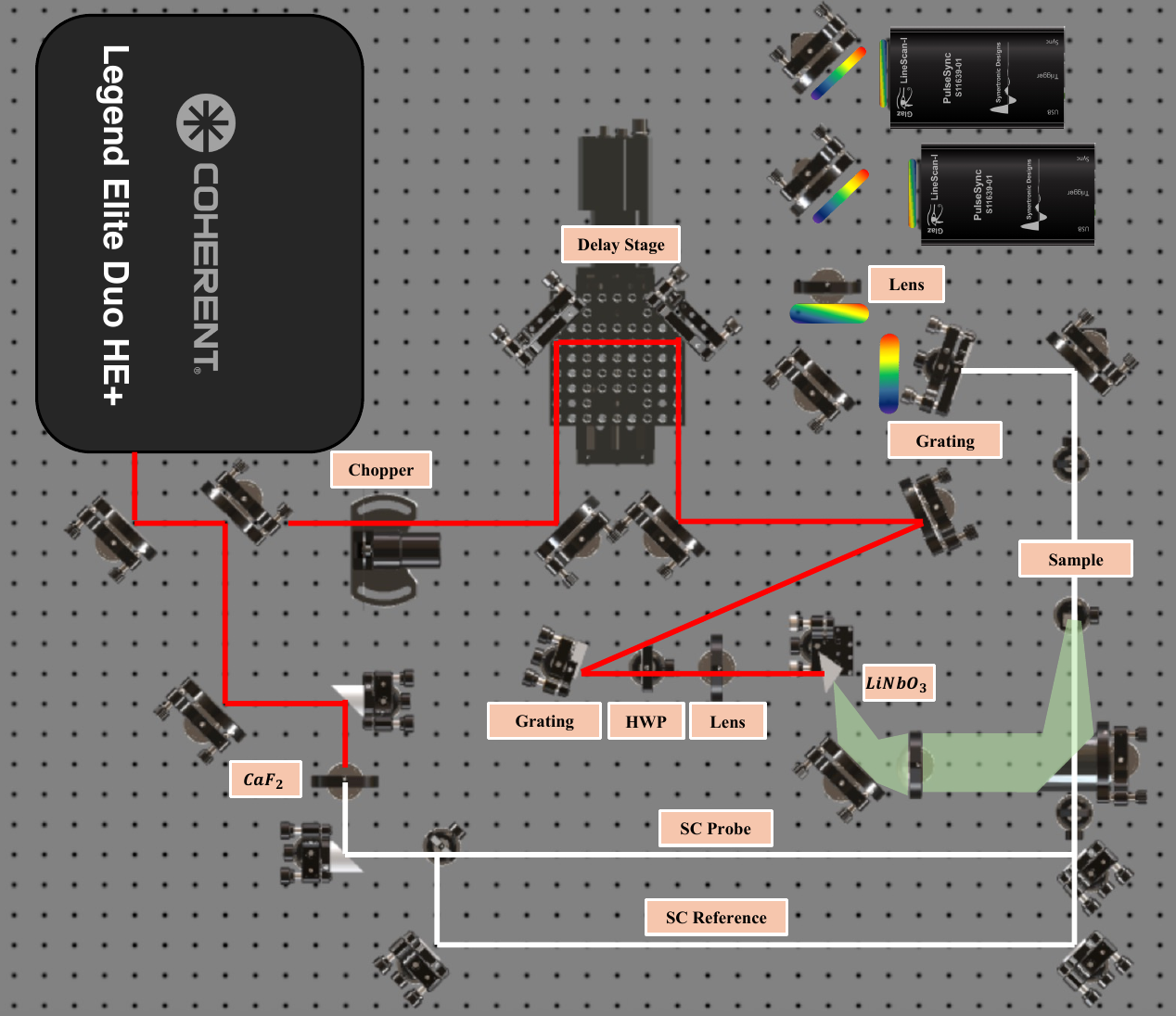}
    \caption{\textbf{Experimental setup for THz Stark spectroscopy with a reference-corrected supercontinuum detection.} Single-cycle THz pulses are produced by tilted-pulse-front excitation in LiNbO$_3$ and are spatially overlapped with femtosecond SC probe pulses generated in CaF$_2$. The probe beam was split before the sample, with one arm passing through the sample while the other serving as a reference to improve signal-to-noise ratio. 
    Both probe and reference spectra are recorded by a spectrometer. A mechanical chopper operating at 500 Hz periodically blocked the THz beam, enabling alternating acquisition of THz-on and THz-off signals for differential detection.}
    \label{fig:Setup}
\end{figure}

\subsubsection{Sample Preparation}
Methyl Orange and Malachite Green were dissolved in deionized water at concentrations of \(5.5~\mathrm{mM}\) and \(9.83~\mathrm{mM}\), respectively, optimized for strong but non-saturating absorption in the visible spectral range.
The solutions were placed in a quartz cuvette with a 100 $\mu m$ path length, ensuring minimal absorption from water itself while maintaining sufficient signal strength. All measurements were performed at room temperature (295 K) to demonstrate TRTSS applicability in liquid-phase systems.

\subsubsection{Data Acquisition and Analysis}
To extract molecular Stark parameters, we performed a Liptay analysis by decomposing the Stark spectra into a linear combination of the absorption spectrum and its derivatives. Following established methods, the derivatives were obtained by direct numerical differentiation of the experimental absorption spectrum. To minimize noise, the data were smoothed with a Savitzky–Golay filter while preserving curvature and peak intensities \cite{Bublitz1997Stark}. As discussed by Bublitz and Boxer, when multiple absorption bands with different electro-optic parameters contribute, a deconvolution approach, fitting the absorption as a sum of band components and using their analytical derivatives, provides an equivalent but species resolved basis \cite{Bublitz1997Stark}. Consistent with this, we used the direct derivative method for both molecules. The extracted Stark parameters, including \(\Delta\mu\), \(\zeta\), and \(\mathrm{Tr}(\Delta\alpha)\), were determined from best fits of the derivative components and compared with equilibrium-solvated TD-DFT values.

\subsubsection{Polarization anisotropy}
\label{SI:Polarization anisotropy}
For an isotropic solution, Liptay theory predicts that the electroabsorption spectra recorded with probe polarization parallel and perpendicular to the THz field have the same linear combination of $A(\nu)$, $A^\prime(\nu)$, and $A^{\prime\prime}(\nu)$, differing only by geometry factors, thus the ratio $R(\lambda)=\Delta\varepsilon_{\parallel}(\lambda)/\Delta\varepsilon_{\perp}(\lambda)$ is wavelength-independent\cite{Bublitz1997Stark,JALVISTE200730}. We evaluate $R(\lambda)$ from the derivative reconstructions within the absorption band and report the band-median with interquartile range (IQR). We obtain, $R_{\mathrm{MG}}=1.75\pm0.28$ (IQR), and $R_{\mathrm{MO}}=1.02\pm0.23$ (IQR), both ratios are flat across the band.\\
\begin{table}[h]
\centering
\caption{Polarization anisotropy metrics from derivative-only diagnostics.}
\begin{tabular}{|l|c|c|}
\hline
Sample & $R_{\mathrm{median}}$ (IQR) & Best scale $s$  \\
\hline
Malachite Green & 1.75 \ (1.62-1.90)     & 1.73 \\
Methyl Orange & 1.02 \ (0.85–1.08) & 0.97  \\
\hline
\end{tabular}
\end{table}
\begin{figure} [H] %[ht!]
    \centering
    \includegraphics[width=1\columnwidth]{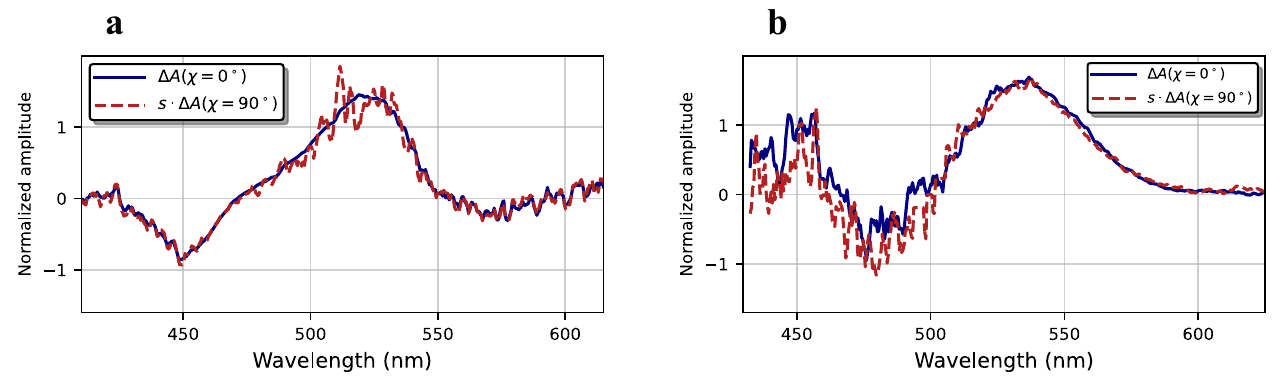}
    \caption{\label{fig:Polarization anisotropy}
    \textbf{a, b}
     Overlay of the time-averaged THz Stark spectrum measured with the probe polarization parallel to the THz electric field (solid blue curve) and the best-scaled spectrum measured with the perpendicular probe polarization (red dashed curve) for Malachite Green and Methyl Orange, respectively.
    }
\end{figure}
To visualize the constraint, we overlay the parallel spectrum $(\chi=0^\circ)$ with the best scaled perpendicular spectrum $s \cdot(\chi=90^\circ)$ in Figure \ref{fig:Polarization anisotropy}.

\subsubsection{Background measurements}
A pure water TRTSS dataset was measured under identical conditions, including cell, path length, THz/probe polarizations, and fluences. The response is broad and nearly wavelength-independent across 400--650 nm, as shown in Figure~\ref{fig:Water spectroscopy}. Since liquid water has no electronic absorption in this spectral window, this signal cannot correspond to a resonant Stark absorption lineshape. The pure-water response is therefore assigned to a nonresonant background rather than to a molecular THz Stark signal. Its broad spectral shape and the absence of derivative-like features are consistent with THz-induced birefringence/Kerr-type effects. The sign reversal observed when the analyzer is rotated by \(90^\circ\) further supports a birefringence-type contribution \cite{Sajadi2017}. Importantly, this does not imply full orientational relaxation of water dipoles during the THz pulse; rather, it reflects an ultrafast field-induced anisotropy of the liquid.

\begin{figure}[H]
    \centering
    \includegraphics[width=0.7\columnwidth]{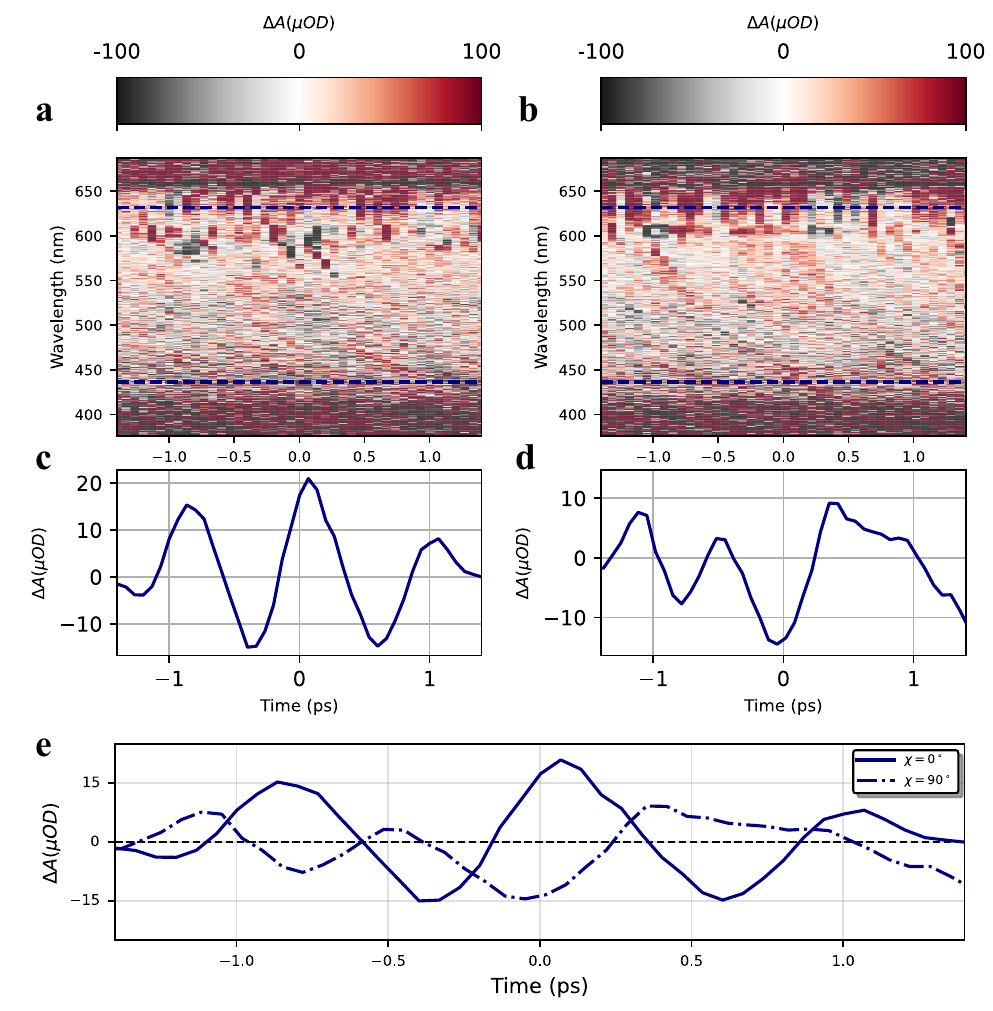}
    \caption{\label{fig:Water spectroscopy}
    \textbf{Pure-water background response.}
    \textbf{a,b} \(\Delta A(\lambda,t)\) maps for parallel and perpendicular geometries, showing a broad, nearly wavelength-independent signal around time zero without derivative-like Stark features.
    \textbf{c,d} Wavelength-averaged \(\Delta A(t)\) traces over the indicated spectral region.
    \textbf{e} Comparison of the averaged traces for the two polarization geometries, showing a sign reversal consistent with a THz-induced birefringence/Kerr-type background.
    }
\end{figure}
Figure~\ref{fig:Water spectroscopy} shows the pure-water background measurements for parallel and perpendicular geometries, together with the corresponding wavelength-averaged traces.

\subsubsection{Local field correction factor}
The electric field experienced by a molecule inside a solvent is different from the macroscopic field applied to the bulk liquid. To account for this difference, we used a local-field correction factor, \(f_L\), which relates the macroscopic THz field to the effective field acting on the solute molecule. This correction describes how the dielectric solvent environment modifies the electric field inside the microscopic cavity occupied by the solute molecule when an external electric field is applied \cite{Aubret2019,Onsager1936,Jackson1998_CED}.

The static dielectric constant of water  \(\varepsilon_s \approx 80\)  cannot be used directly at THz frequencies, since the collective reorientation of water dipoles occurs on a timescale of \(8.3~\mathrm{ps}\) \cite{Arbe2016WaterDielectricSusceptibility} and cannot follow a sub-picosecond THz field. The complex permittivity of water in the THz range, \(\varepsilon(\omega)\), was therefore evaluated using the model of Liebe, Hufford, and Manabe \cite{Liebe1991WaterPermittivity}.

The local-field correction was calculated in analogy to the ellipsoidal-cavity model used previously for THz Stark spectroscopy \cite{kang2024,Aubret2019,Onsager1936}. Malachite Green and Methyl Orange were approximated as occupying ellipsoidal molecular cavities in water. For each principal molecular axis, a local-field factor was calculated from the frequency-dependent dielectric response of water and the corresponding depolarization factor of the ellipsoid. The scalar local-field factor was then obtained by averaging the three principal-axis contributions.

Because the THz pulse is broadband, the local-field factor is formally frequency-dependent. We therefore evaluated \(f_L(\omega)\) over the measured THz spectral bandwidth (0.258 to 0.886 THz) and calculated a spectrum-weighted effective local-field factor for each molecule. Since \(f_L(\omega)\) varied only weakly over the spectral region carrying the main THz intensity, these effective values were used as constant correction factors in the extraction of the Stark parameters. Over this spectral region, the variation of \(f_L(\omega)\) was below 5.88\% for Malachite Green and 5.71\% for Methyl Orange. 
The resulting effective correction factors were \(f_L^\mathrm{eff}=1.40\) for Malachite Green and \(f_L^\mathrm{eff}=1.38\) for Methyl Orange. The extracted dipole-moment changes were corrected by dividing \(\Delta\mu\) by \(f_L^\mathrm{eff}\), while the polarizability changes were corrected by dividing \(\mathrm{Tr}(\Delta\alpha)\) by \((f_L^\mathrm{eff})^2\). This correction accounts for the average dielectric screening of the THz field by water.

\subsubsection{Effective THz field strength in the water cuvette}
The effective field strength at which the probe pulse interrogates the molecular system is smaller than the incident free-space THz field because of several effects, including Fresnel reflections and Fabry--Pérot interferences at the cuvette interfaces, THz absorption by water, finite sample thickness, finite probe-pulse duration, and group-velocity mismatch between the THz waveform and the optical probe pulse. To estimate this reduction, finite-difference time-domain simulations were used for the THz waveform propagating through the cuvette and the \(100~\mu\mathrm{m}\) water layer, in analogy to our previous THz Stark analysis \cite{kang2024}.

The simulated field was post-processed to obtain the effective field sampled by the time-delayed optical probe inside the water layer. The calculation gives an effective-field reduction factor of approximately \(0.35\) for the \(100~\mu\mathrm{m}\) water cuvette, such that $
E_\mathrm{eff} \approx 0.35 E_\mathrm{air},$
where \(E_\mathrm{air}\) is the incident peak THz field at the sample position in air. The absolute value of \(E_\mathrm{eff}\) therefore depends on the experimental field calibration, while the factor above describes the propagation and probe-averaging reduction for the simulated cuvette geometry.

\subsubsection{Computational Methods}
Electronic structure calculations were performed with Gaussian 16 software. First, the ground-state geometries were optimized using a DFT description with B3LYP functional and 6-31+G(d,p) basis set. The counterions are expected to be solvated and stabilized in a hihgly polar solvent such as water, and therefore assumed to remain sufficiently far from the dye molecules. Under these conditions, the system can be treated as individual solvated molecules without explicitly including counterions. Therefore, the charge of Malachite Green was set to +1 and the charge of Methyl Orange to -1. The ground and excited state properties including electronic dipole moment, polarizabilities, transition dipole moment and energies were computed with the range-separated wB97XD functional and the same basis set as for the geometry optimization. Solvent effects were included within the polarizable continuum model (PCM) both for the conformational analysis and for the calculation of excitation properties. The ability of the selected functional to describe the mixed charge transfer (CT) and locally excited (LE) states was validated by comparing the vacuum properties with the CC2 approach with a def2-TZVPPD basis set, which was shown to provide similar precision for both the CT and LE states (Fig. S\ref{fig:CC2_vs_TDDFT} and Table S\ref{tab:CC2_vs_TDDFT}). In the next step, we investigated the solvent effects on the electronic properties by comparison of the gas phase properties with the nonequilibrium linear response solvation, including only fast response of the electronic degrees of freedom, and the equilibrium solvation including nuclear relaxation effects (Table S\ref{tab:Solv_effects}). In order to validate the computed excited-state properties and visualize the effects of the environment, we computed absorption spectra for the two studied systems and three different solvation schemes and compared them with experiment. 

\begin{figure} [H] %[ht!]
    \centering
    \includegraphics[width=1.0\columnwidth]{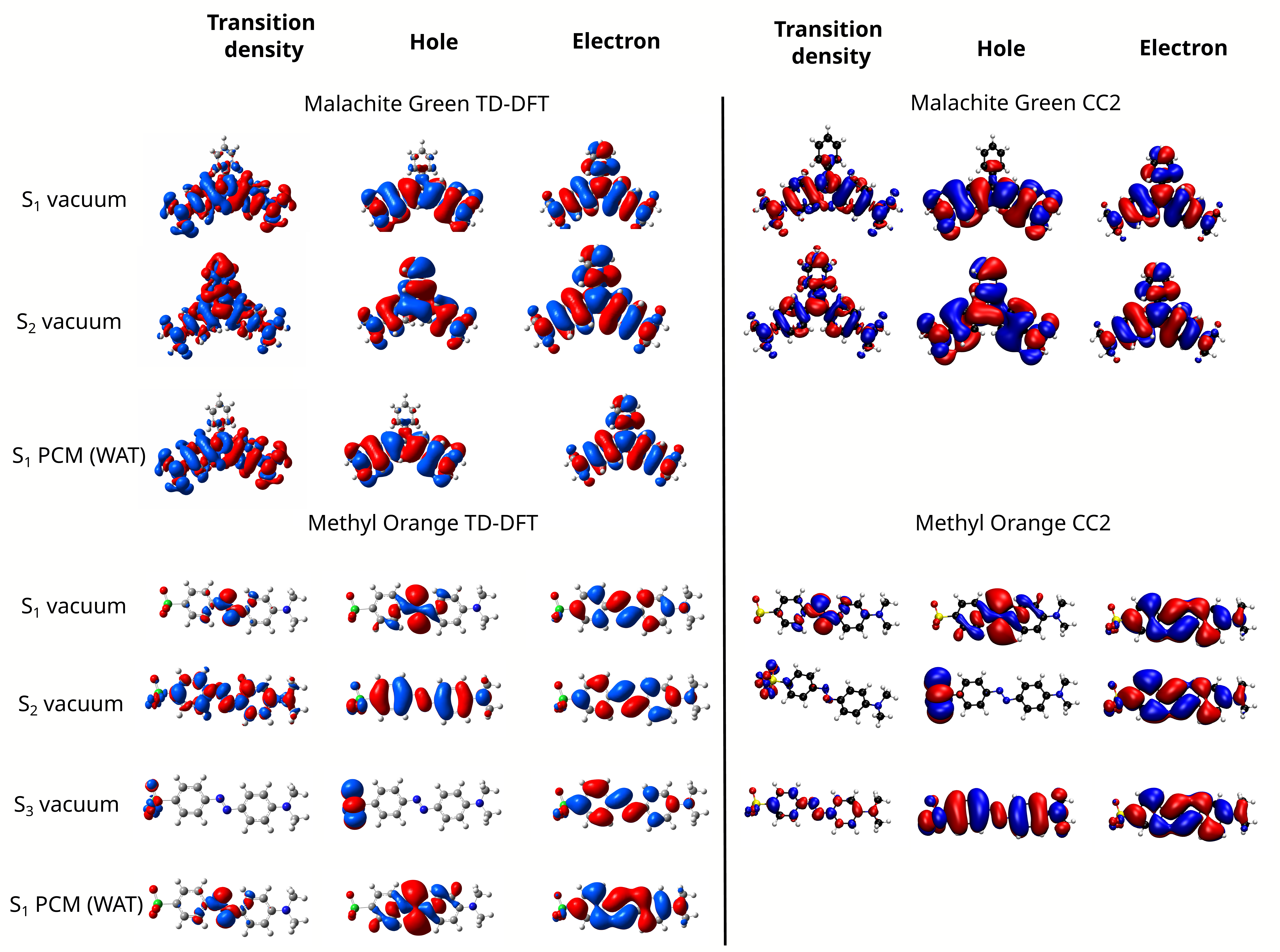}
    \caption{\label{fig:CC2_vs_TDDFT} Comparison of the natural transition orbitals and transition densities between TD-DFT with wB97XD DFT functional with 6-31+G(d,p) basis and CC2 calculation with def2-TZVPPD in vacuum. TD-DFT orbitals and densities results agree well with the CC2 results except for the Methyl Orange second and third excited states that are interchanged. The natural transition orbitals keep the same shape also in water solution modeled as polarizable continuum environment.}
\end{figure}

\begin{figure} [H] %[ht!]
    \centering
    \includegraphics[width=0.8\columnwidth]{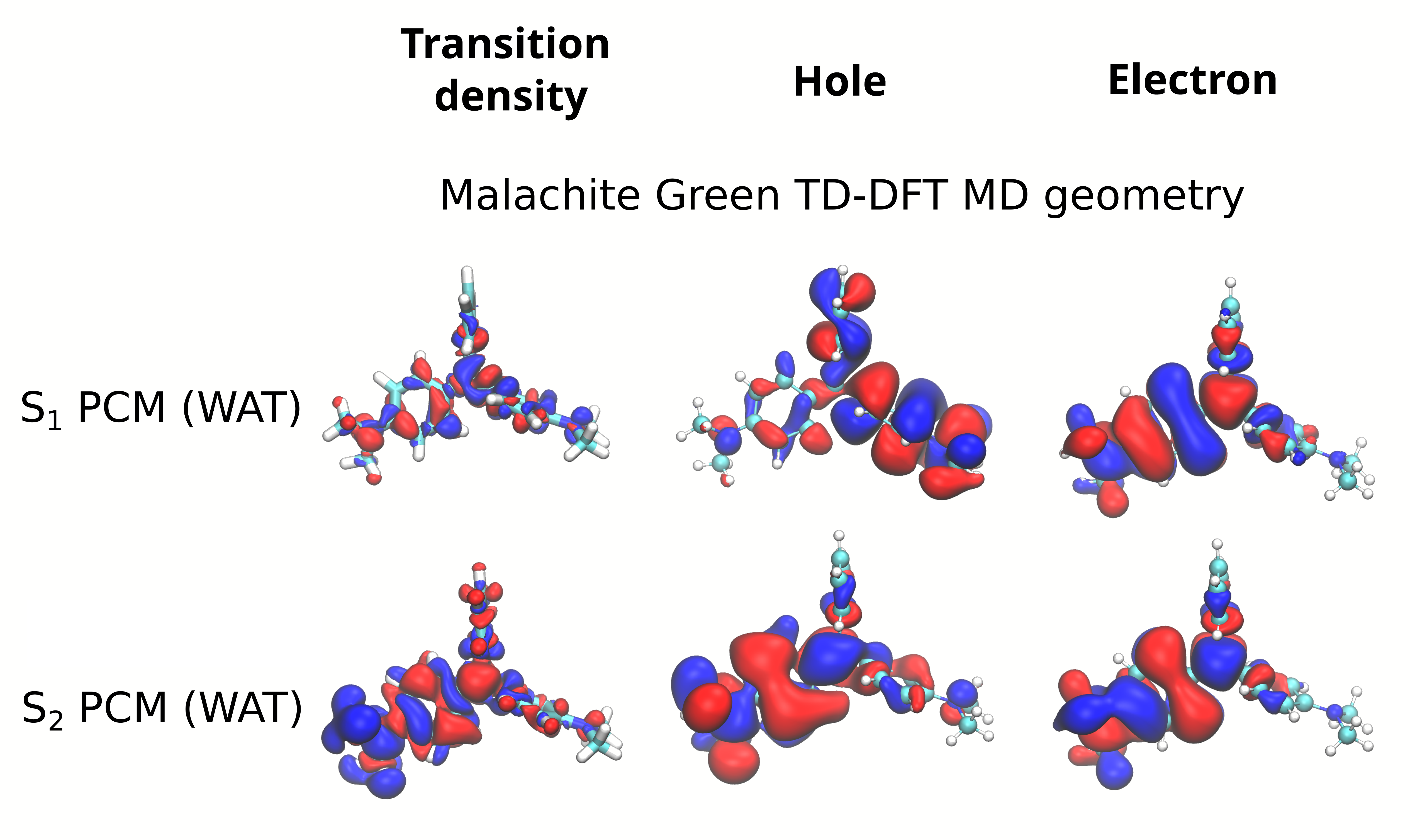}
    \caption{\label{fig:MG_MD_geometry} The Natural transition orbitals and transition densities computed with TD-DFT with wB97XD DFT functional with 6-31+G(d,p) basis for a typical structure obtained from classical MD. For the distorted structure resulting from thermal fluctuations, the molecular orbitals are more localized on individual dimethylaminophenyl groups. This localization is caused by rotation of the dimethylaminophenyl units and reduced \(\pi\)-electron overlap between the two units.}
\end{figure}

\begin{table}[H] %[ht!]
\begin{tabular}{ |p{3.0cm}|p{3.5cm}|p{3.5cm}|p{3.5cm}| }
\hline
Property & wB97XD (B3LYP geom) &  CC2 (B3LYP geom)  & CC2 (CC2 geom) \\
\hline
\multicolumn{4}{|c|}{Malachite green} \\
\hline
$E_{ge}$ (eV)   &  2.766 &   2.236 &   2.232  \\
$\boldsymbol{\mu}_{ge}$ (D)   & (9.66, 0.0, 0.05) & (11.27, 0.0, 0.00) & (11.20, 0.0, 0.00) \\
$\Delta\boldsymbol{\mu}$ (D)   &  (0.0, -2.47, 0.0) &   (0.0, -2.29, 0.0) & (0.0, -2.42, 0.0) \\
\hline
\hline
\multicolumn{4}{|c|}{Methyl orange} \\
\hline
$E_{ge}$ (eV)   &  3.609 &   3.423 &   3.236  \\
$\boldsymbol{\mu}_{ge}$ (D)   & (9.12, 0.0, 0.0) & (9.69, 0.0, 0.00) & (9.86, 0.0, 0.00) \\
$\Delta\boldsymbol{\mu}$ (D)   &  (-2.21, -0.10, 0.05) &   (5.53, -0.20, -0.10) & (-6.24, -0.25, -0.03) \\

\hline
\end{tabular}
\caption{\label{tab:CC2_vs_TDDFT} Comparison of the electronic properties for the first excited state between the TD-DFT and CC2. \(E_{ge}\) and \(\boldsymbol{\mu}_{ge}\) correspond to the vertical excitation energy and transition dipole moment components, respectively, between the ground state and first excited state \(S_1\). \(\Delta\boldsymbol{\mu}\) represents the vector difference between excited- and ground-state dipole moments.}
\end{table}

\begin{table}[H] %[ht!]
\begin{tabular}{ |p{3.0cm}|p{3.5cm}|p{4.0cm}|p{3.5cm}| }
\hline
Property & Gas phase &  Non-equilibrium Solv.  & Equilibrium Solv. \\
\hline
\multicolumn{4}{|c|}{Malachite green} \\
\hline
$E_{ge}$ (eV)   &  2.766 &   2.603 &   2.359  \\
$\boldsymbol{\mu}_{ge}$ (D)   & (9.66, 0.0, 0.05) & (10.65, 0.0, -0.05) & (12.35, 0.0, 0.08) \\
$\Delta\boldsymbol{\mu}$ (D)   &  (0.0, -2.47, 0.0) &   (0.0, -3.72, 0.0) & (0.0, -4.17, 0.0) \\
Tr($\Delta\boldsymbol{\alpha}_{diag}$) (\AA$^3$)   & (8.9, 0.8, 0.5)  & (31.7, -0.7, 0.7)  & (31.8, -0.7, 0.7) \\
\hline
\hline
\multicolumn{4}{|c|}{Methyl orange} \\
\hline
$E_{ge}$ (eV)   &  3.609 &   3.364 &   3.191  \\
$\boldsymbol{\mu}_{ge}$ (D)   & (9.12, 0.0, 0.0) & (9.68, 0.04, 0.08) & (10.58, -0.05, 0.10) \\
$\Delta\boldsymbol{\mu}$ (D)   &  (-2.21, -0.10, 0.05) &   (-8.16, 0.25, -0.05) & (-8.70, 0.23, -0.05) \\
Tr($\Delta\boldsymbol{\alpha}_{diag}$) (\AA$^3$)   & (103.4, 0.6, 1.0)  & (82.4, 0.6, 0.7)  & (90.6, 0.4, 0.7) \\
\hline
\end{tabular}
\caption{\label{tab:Solv_effects} Comparison of the electronic properties for the different solvation scheme for TD-DFT with wB97XD and 6-31+G(d,p) basis set.}
\end{table}

\end{suppinfo}

\newpage

\bibliography{acs-achemso}

\end{document}